\documentclass[prb,amsmath,amssymb,superscriptaddress,twocolumn]{revtex4}
\usepackage{float}
\usepackage{amsmath}
\usepackage{amssymb}
\usepackage{amsfonts}
\usepackage{euscript}
\usepackage{enumerate}
\usepackage{hhline}
\usepackage{pslatex}
\usepackage{tabularx}
\usepackage[usenames,dvipsnames]{xcolor}

\usepackage{graphicx}

\usepackage{dcolumn}
\usepackage{bm}
\usepackage[sort&compress]{natbib}

\usepackage{lipsum}

\makeatletter
\renewcommand{\@biblabel}[1]{#1. }
\renewcommand{\@dotsep}{500}
\renewcommand{\@pnumwidth}{0em}
\renewcommand{\l@figure}[2]{
\@dottedtocline{1}{1.5em}{2em}{Figure #1}{}\vspace{15pt}}

\usepackage[normalem]{ulem}

\begin{document}
\title{Rod and slit photonic crystal microrings for on-chip cavity quantum electrodynamics}

\author{Xiyuan Lu}\email{xiyuan.lu@nist.gov}
\affiliation{Microsystems and Nanotechnology Division, Physical Measurement Laboratory, National Institute of Standards and Technology, Gaithersburg, MD 20899, USA}
\affiliation{Joint Quantum Institute, NIST/University of Maryland, College Park, MD 20742, USA}
\author{Feng Zhou} 
\affiliation{Microsystems and Nanotechnology Division, Physical Measurement Laboratory, National Institute of Standards and Technology, Gaithersburg, MD 20899, USA}
\affiliation{Joint Quantum Institute, NIST/University of Maryland, College Park, MD 20742, USA}
\author{Yi Sun}
\affiliation{Microsystems and Nanotechnology Division, Physical Measurement Laboratory, National Institute of Standards and Technology, Gaithersburg, MD 20899, USA}
\affiliation{Joint Quantum Institute, NIST/University of Maryland, College Park, MD 20742, USA}
\author{Mingkang Wang} 
\affiliation{Microsystems and Nanotechnology Division, Physical Measurement Laboratory, National Institute of Standards and Technology, Gaithersburg, MD 20899, USA}
\affiliation{Department of
Chemistry and Biochemistry, University of Maryland, College Park, Maryland 20742, USA}
\author{Qingyang Yan}
\affiliation{MicroKerr Photonics, 20404 Highland Hall Drive, MD 20886, USA}
\author{Ashish Chanana} 
\affiliation{Microsystems and Nanotechnology Division, Physical Measurement Laboratory, National Institute of Standards and Technology, Gaithersburg, MD 20899, USA}
\author{Andrew McClung}
\affiliation{Department of Electrical and Computer Engineering, University of Massachusetts Amherst, Amherst, MA 01003, USA}
\author{Vladimir A. Aksyuk}
\affiliation{Microsystems and Nanotechnology Division, Physical Measurement Laboratory, National Institute of Standards and Technology, Gaithersburg, MD 20899, USA}
\author{Marcelo Davanco}
\affiliation{Microsystems and Nanotechnology Division, Physical Measurement Laboratory, National Institute of Standards and Technology, Gaithersburg, MD 20899, USA}
\author{Kartik Srinivasan} \email{kartik.srinivasan@nist.gov}
\affiliation{Microsystems and Nanotechnology Division, Physical Measurement Laboratory, National Institute of Standards and Technology, Gaithersburg, MD 20899, USA}
\affiliation{Joint Quantum Institute, NIST/University of Maryland, College Park, MD 20742, USA}

\date{\today}

\begin{abstract}
\noindent Micro-/nanocavities that combine high quality factor ($Q$) and small mode volume ($V$) have been used to enhance light-matter interactions for cavity quantum electrodynamics (cQED). Whispering gallery mode (WGM) geometries such as microdisks and microrings support high-$Q$ and are design- and fabrication-friendly, but $V$ is often limited to tens of cubic wavelengths to avoid WGM radiation. The stronger modal confinement provided by either one-dimensional or two-dimensional photonic crystal defect geometries can yield sub-cubic-wavelength $V$, yet the requirements on precise design and dimensional control are typically much more stringent to ensure high-$Q$. Given their complementary features, there has been sustained interest in geometries that combine the advantages of WGM and photonic crystal cavities. Recently, a `microgear' photonic crystal ring (MPhCR) has shown promise in enabling additional defect localization ($>$ 10$\times$ reduction of $V$) of a WGM, while maintaining high-$Q$ ($\approx10^6$) and other WGM characteristics in ease of coupling and design. However, the unit cell geometry used is unlike traditional PhC cavities, and etched surfaces may be too close to embedded quantum nodes (quantum dots, atomic defect spins, etc.) for cQED applications. Here, we report two novel PhCR designs with `rod' and `slit' unit cells, whose geometries are more traditional and suitable for solid-state cQED. Both rod and slit PhCRs have high-$Q$ ($>10^6$) with WGM coupling properties preserved. A further $\approx$~10$\times$ reduction of $V$ by defect localization is observed in rod PhCRs. Moreover, both fundamental and 2nd-order PhC modes co-exist in slit PhCRs with high $Q$s and good coupling. Our work showcases that high-$Q/V$ PhCRs are in general straightforward to design and fabricate and are a promising platform to explore for cQED.
\end{abstract}  

\maketitle
\section{Introduction}
\vspace{-0.155in}
\noindent Optical micro-/nanocavities enhance light-matter interactions through both enhanced photon lifetime (proportional to the cavity quality factor $Q$) and strong spatial confinement (quantified by the effective mode volume $V$)~\cite{Vahala_Nature_2003_Optical}, with whispering gallery mode (WGM) and defect photonic crystal (dPhC) microcavities being two commonly utilized platforms in integrated photonics. WGM microcavities support high-$Q$ (typically $\gtrsim$10$^6$) and moderate $V$ (typically tens of cubic wavelengths), and are straightforward with respect to design and fabrication. $V$ is mainly restricted by the radius needing to be large enough to support total internal reflection at the curved interface. External coupling, typically achieved through evanescent mode coupling to an adjacent waveguide, is mature and advanced coupling tasks can be achieved, for example, by pulley couplers~\cite{Li_NatPhoton_2016_Efficient, Lu_NatPhys_2019_Chip}. In comparison, through proper design and fabrication, dPhC cavities can support sub-cubic-wavelength $V$ with high $Q$ similar to WGMs, and they are thus often considered an optimal choice for cavity quantum electrodynamics (cQED)~\cite{lodahl_interfacing_2015,janitz_cavity_2020}. Though recipes such as momentum space design~\cite{Srinivasan_OE_2002_Momentum}, inverse problem approach~\cite{Englund_OE_2005_General}, and other deterministic methods~\cite{Quan_OE_2011_Deterministic,Lalanne_LPR_2008_Photon} exist for dPhC design, optimization for simultaneous high-$Q$ and small-$V$ and subsequent fabrication with a pattern fidelity adequate for retaining those simulated properties can be time-consuming and resource intensive.

\begin{figure*}[t!]
\centering\includegraphics[width=0.88\linewidth]{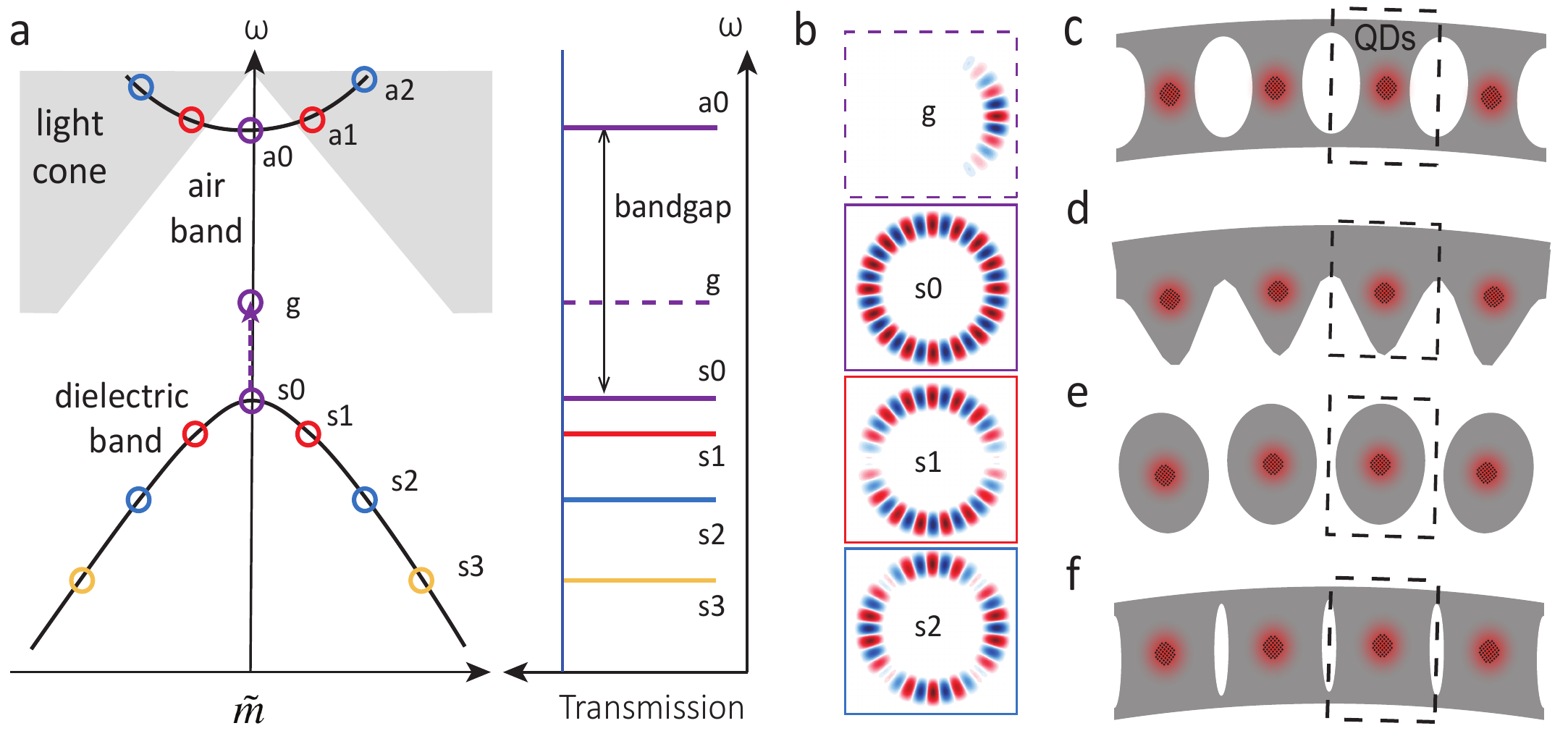}
\caption{\textbf{Photonic crystal ring (PhCR) for cavity quantum electrodynamics (cQED).} \textbf{a,} Band diagram for a PhCR. Air band modes are pushed up in frequency towards the light cone, and only those below the light cone are observed in waveguide-coupled transmission measurements. Dielectric band modes are compressed in free spectral range (the frequency separation between adjacent modes) around the band edge. The frequency difference of the air and dielectric band-edge modes $a0$ and $s0$ is the bandgap. Through defect localization, a defect mode $g$ can be created from $s0$, with its frequency pushed into the bandgap. These mode characteristics can be observed in cavity transmission spectroscopy if the modes are properly waveguide-coupled. \textbf{b,} Illustration of the spatial electric field profiles of the defect and dielectric band-edge modes. All these modes are a redistribution in azimuthal angle, with similar mode profiles in a unit cell. \textbf{c-f,} Illustration of four types of unit cells in PhCRs, whose characteristic shapes are referred to hereafter as `holes', `microgear', `rods', and `slits', respectively. Quantum dots (QDs), as a representative quantum emitter for future cQED experiments, are illustrated at the center positions of each cell.}
\label{Fig1}
\end{figure*}

Given their complementary advantageous features, there have thus been long-standing efforts to combine WGM and PhC microcavities~\cite{Smith_APL_2001_Coupled, Kim_APL_2002_Two, Zhang_OL_2014_High,Lee_OL_2012_Slow, Gao_SciRep_2016_Air, KML_OE_2014_Maximizing,Lu_NatPhoton_2022_High,Lucas_arXiv_2022_Tailoring}. Most efforts come from two paths. The first is to create polygonal or circular line defects in two-dimensional PhCs, referred as photonic crystal ‘disk/ring’ resonators (or PCDRs/PCRRs, respectively)~\cite{Smith_APL_2001_Coupled, Kim_APL_2002_Two, Zhang_OL_2014_High}. This method naturally retains the small $V$s of PhCs, but the coupling is more similar to PhCs than WGM resonators, and $Q$s have been limited to $<10^5$. The second approach is to implement a PhC cell in a microring [Fig.~\ref{Fig1}(a,b)], for example, by drilling holes in the ring~\cite{Lee_OL_2012_Slow, Gao_SciRep_2016_Air, KML_OE_2014_Maximizing}, as illustrated in Fig.~\ref{Fig1}(c). This method retains the mode and coupling properties of typical WGMs, but with lower $Q$ ($<10^5$), as well as  much larger mode volumes than dPhC resonators. The band diagram is well understood and often characterized in cavity transmission, as illustrated by Fig.~\ref{Fig1}(a,b). To address the limitations of these previous approaches, recently, a `microgear' photonic crystal ring (MPhCR), as illustrated in Fig.~\ref{Fig1}(d), has been reported~\cite{Lu_NatPhoton_2022_High,Lucas_arXiv_2022_Tailoring}, with origins derived from at least two independent precursors, PhC microcavities in the perturbative regime~\cite{Lu_APL_2014_Selective} and `alligator' photonic crystal waveguides~\cite{Yu_APL_2014_Nanowire}. Implemented in silicon nitride~\cite{Lu_NatPhoton_2022_High} and tantala~\cite{Lucas_arXiv_2022_Tailoring} thin-film photonics platforms, both designs support band-edge states with high $Q\approx10^6$, similar to conventional WGM microrings. Moreover, the MPhCR~\cite{Lu_NatPhoton_2022_High} has shown promise in enabling additional defect localization ($>10\times$ reduction of $V$) of the band-edge WGMs, while maintaining $Q\approx10^6$, WGM characteristics in coupling, and straightforward design without detailed optimization. This MPhCR geometry is suitable for single mode lasing~\cite{Feng_Science_2014_Single, Arbabi_OE_2015_Grating} and wide-band nonlinear optical processes~\cite{Moss_NatPhoton_2013_New}, including frequency comb generation~\cite{Yu_NatPhoton_2021_Spontaneous, Lucas_arXiv_2022_Tailoring,moille_DEMS_preprint}.

However, the unit cell in the MPhCR is quite different from typical PhC designs, calling to question whether this specific geometry is the only one that can yield high-$Q$ in a PhCR. In particular, many other PhCR geometries have been studied, including fully-etched holes~\cite{Lee_OL_2012_Slow}, partially-etched holes~\cite{Urbonas_OL_2016_Air}, and circular `rods'~\cite{Timmerman_OE_2022_Nanorod}, but none has been experimentally demonstrated with high $Q$ close to that of the MPhCR -- typical $Q$s have been at least 10$\times$ lower. In addition, a particular problem of the MPhCR for quantum optics applications is that its etched surfaces may be too close to quantum emitters (quantum dots, atomic defect spins, etc.), as illustrated in Fig.~\ref{Fig1}(d), which can create traps and surface states that lead to spectral diffusion and dephasing and thus degrade the quantum emitter's coherence~\cite{Liu_PRAppl_2018_Single}. It would thus be beneficial to separate the etched surfaces as far from the quantum emitters as possible.

In this paper, we report two types of PhCR designs with `rod' and `slit' unit cells whose geometries are similar to traditional PhC unit cells (as illustrated in Fig.~\ref{Fig1}(e,f)). These designs also allow more space between etched surfaces and potential integrated quantum emitters. Working in the silicon nitride platform, both rod and slit PhCRs have band-edge WGMs with $Q>10^6$ and waveguide coupling properties preserved in comparison to standard WGMs. A further $\approx$~10$\times$ reduction of $V$ by defect localization is observed experimentally in `rod' PhCRs. Moreover, both fundamental and 2nd-order PhC band-edge modes co-exist in `slit' PhCRs with high $Q$s and good coupling at the same time.

Our work showcases that high-$Q/V$ PhCRs in various geometries are straightforward to design and fabricate, and highlights the platform's promise for applications in quantum optics. In particular, in the context of quantum optics with single quantum emitters (including cQED), defects in 2D materials~\cite{peyskens_integration_2019,parto_cavity-enhanced_2022}, single organic molecules~\cite{toninelli_single_2021}, and colloidal quantum dots~\cite{chen_deterministic_2018,elsinger_integration_2019} have all been integrated with silicon nitride photonics. Moreover, there has been recent work investigating single quantum emitters that are directly hosted in silicon nitride~\cite{senichev_room-temperature_2021}. Finally, our PhCR approach can be translated to other single quantum emitter platforms, by adapting our designs for those materials or through heterogeneous integration~\cite{Davanco_NatCommun_2017_Heterogeneous}.  

\begin{figure*}[t!]
\centering\includegraphics[width=0.88\linewidth]{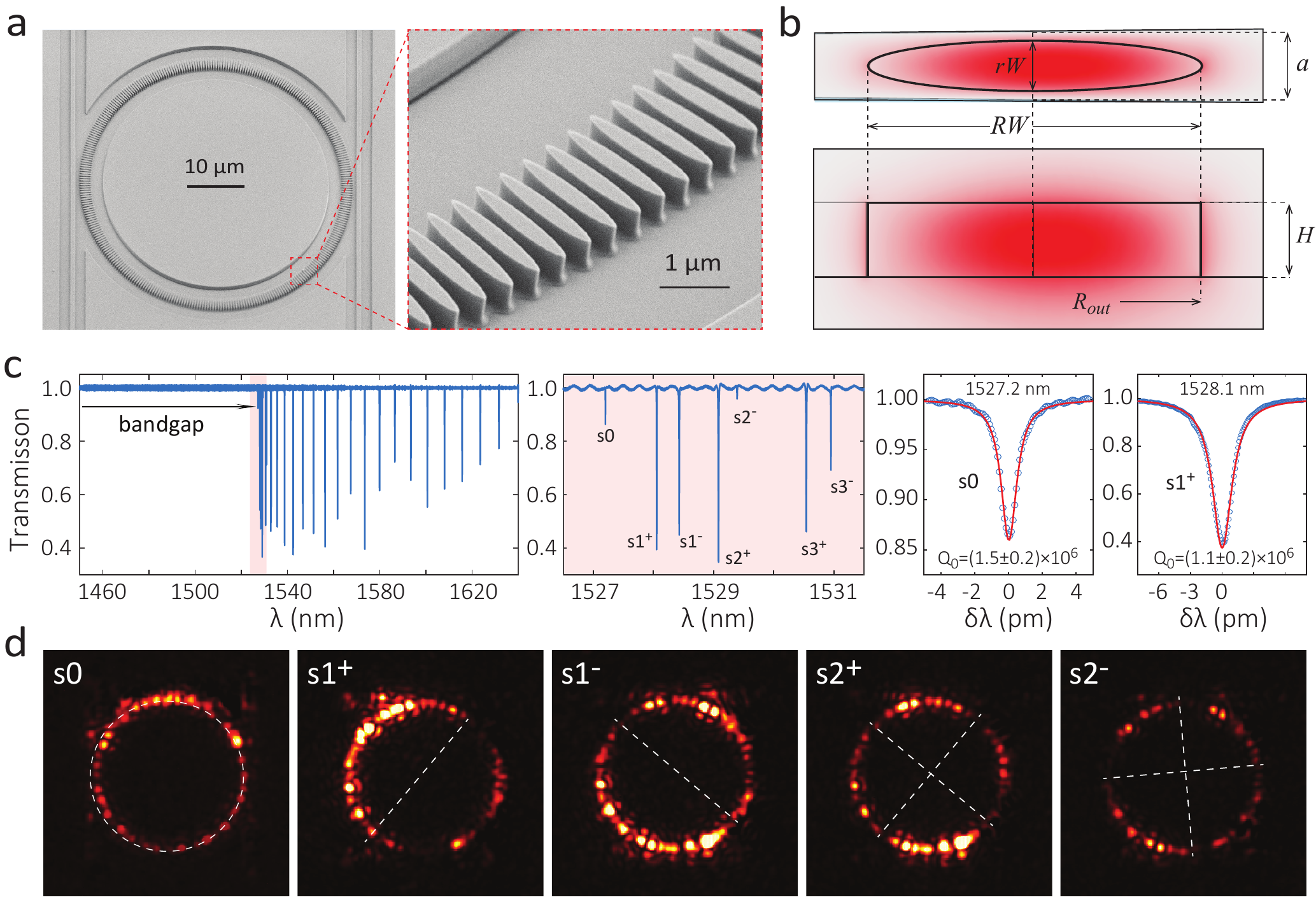}
\caption{\textbf{Introducing the rod photonic crystal ring (rPhCR).} \textbf{a,} SEM images of the rPhCR device and zoom-in view of several units cells of the rod structure. The silicon nitride rods sit on top of a silicon dioxide substrate with top air cladding. \textbf{b,} A finite-element-method simulation of a unit cell of the rPhCR device, showing the dominant (radial-direction) electric field profile in top and cross-section views. The cell is 1/324 of the rPhCR circumference, with a cell width of $a=$~460~nm and a thickness of $H=$~500~nm. The elliptical rod has a major axis length of $RW=$~2.25~$\mu$m and minor axis length of $rW=$~354~nm, and its outer ring radius is $R_{out}=$~24.875~$\mu$m. The simulated resonance frequency for the dielectric band-edge mode ($s0$) is 196.3~THz. \textbf{c,} Transmission spectrum of the rPhCR device that has the same parameters in design as used in simulation, except a rod width of $rW=$~414~nm in the designed pattern. This $rW$ is a critical parameter, and typically has an offset of approximately 60~nm after fabrication. The modes at the dielectric band edge (red area) are displayed in the middle panel, with the modes labeled as $\{s0,s1^{\pm},s2^{\pm},s3^{\pm}\}$, where the labels follow Fig.~\ref{Fig1}(a) and the plus and minus sign indicate a higher and lower frequency, respectively. The two right panels show optical fitting to the $s0$ and $s1^+$ modes, respectively, with optical intrinsic quality ($Q_0$) above 10$^6$. The uncertainties of the $Q_0$ values specified here represent the 95~\% confidence interval of the nonlinear least squares fitting. \textbf{d,} Infrared images of the five band-edge modes $\{s0,s1^{\pm},s2^{\pm}\}$.}
\label{Fig2}
\end{figure*}

\vspace{-0.25in}
\section{`Rod' photonic crystal microring}
\vspace{-0.15in}
\noindent In this section we introduce `rod' PhCRs in the stoichiometric silicon nitride (Si$_3$N$_4$) integrated photonics platform. Apart from the aforementioned experiments with single quantum emitters, this platform has been previously successful for wide-band nonlinear optics~\cite{Moss_NatPhoton_2013_New}, including optical parametric oscillation~\cite{Lu_Optica_2019_Milliwatt} and frequency combs~\cite{Gaeta_NatPhoton_2019_Photonic}. Moreover, heterogeneous integration of lasers, amplifiers, photodetectors, and modulators at sub-micron wavelengths has been achieved onto the silicon nitride photonics platform~\cite{Tran_Nature_2022_Extending}. Our rod PhCR (rPhCR) devices were fabricated 
according to our previous method~\cite{Lu_NatPhoton_2022_High}. 

\begin{figure*}[t!]
\centering\includegraphics[width=0.88\linewidth]{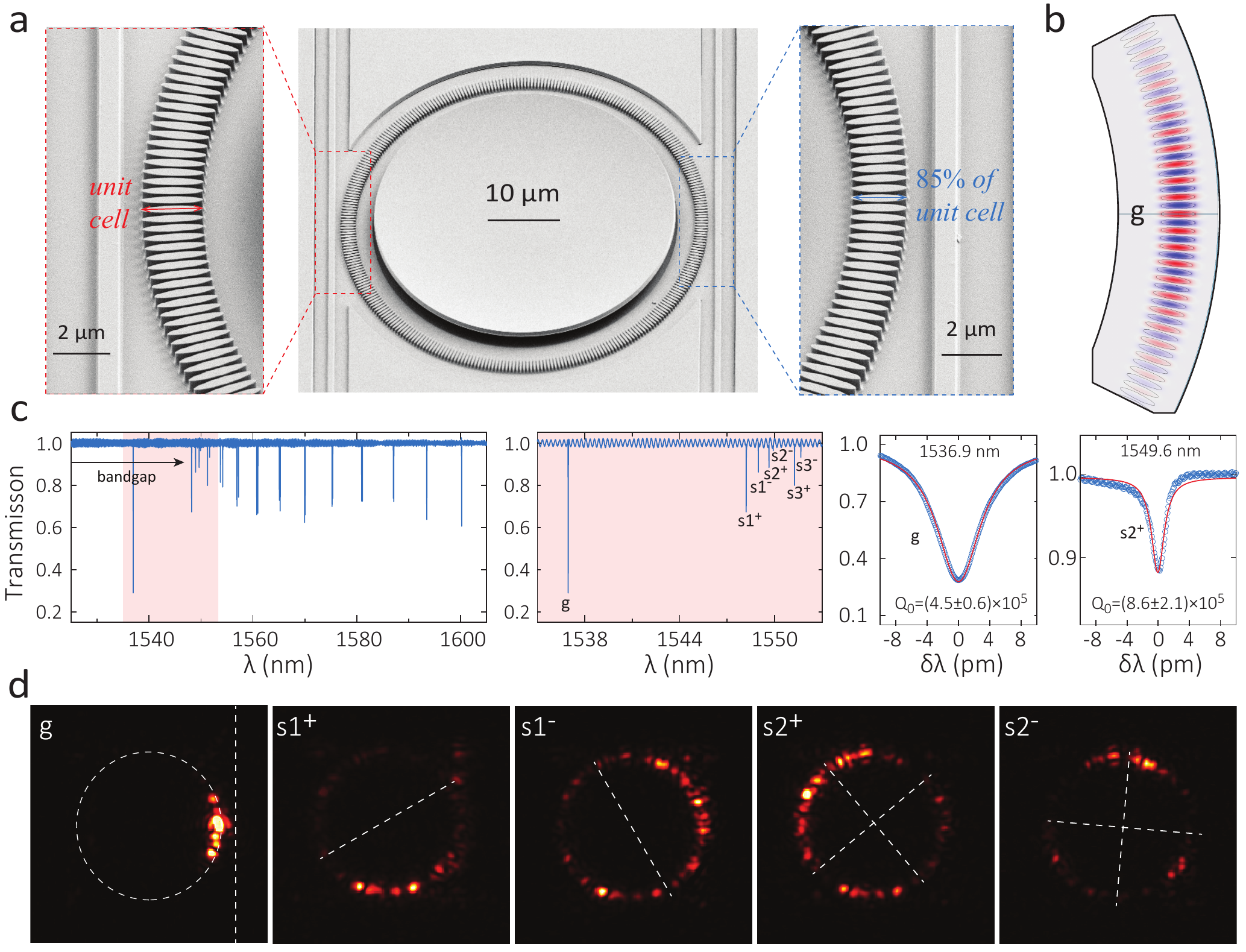}
\caption{\textbf{Implementing defect photonic crystal cavity in the rPhCR.} \textbf{a,} SEM images of a rPhCR device with zoom-in views of cells in the uniform region (left) and the defect region (right). The unit cell parameters are similar to those in Fig.~\ref{Fig2}, except that its outer ring radius is now $R_{out}=$~25~$\mu$m. In the defect region, a scaling of the unit cell is introduced with a quadratic function across 48 cells and a maximal 15~\% reduction of the $RW$ when it is closest to the waveguide. The center of the rods remain fixed in distance to the rPhCR center. \textbf{b,} Simulation of the defect photonic crystal cavity mode, labeled as the $g$ mode hereafter. \textbf{c,} Transmission spectrum of this defect rPhCR device, showing the $g$ mode, dielectric band-edge modes, and the bandgap (which extends as far as the non-defect devices in Fig.~\ref{Fig2} beyond the laser scan range). A zoom-in of the modes at the dielectric band-edge (pale red shaded area) is displayed in the middle panel, labeled as $\{g,s1^{\pm},s2^{\pm}\}$. The right panels show nonlinear least squares fits to the $g$ and $s2^+$ modes, respectively. The $g$ mode has a loaded optical quality of $Q_t =$~(2.4$\pm$0.1)$\times10^5$, with the intrinsic $Q_0$ as displayed. The uncertainty of the $Q$ values represents the 95~\% confidence range of the nonlinear fitting. \textbf{d,} Infrared images of the $\{g,s1^{\pm},s2^{\pm}\}$ modes. This $g$ mode is a localization of the $s0$ mode in a defect-free rPhCR.}
\label{Fig3}
\end{figure*}

We show in Fig.~\ref{Fig2}(a) a scanning electron microcope (SEM) image with the rPhCR with two coupling waveguides. The outer radius of the microring ($R_{out}$) is 25~$\mu$m, containing 324 identical unit cells, with adjacent cells spaced by $a$~=~460~nm. In the measurements we will show, only the right waveguide is used, with a ring-waveguide gap of approximately 650~nm. The zoomed-in image [Fig.~\ref{Fig2}(b)] shows $\approx$~15 cells, with each cell have an an elliptical shape whose major and minor axis lengths are $RW=$~2.25~$\mu$m and $rW=$~354~nm, respectively. We carry out finite-element method simulation of a unit cell, and find the fundamental transverse-electric-like (TE1) mode, $s0$, at approximately 196.3~THz, using refractive indices of 1.98 for the silicon nitride core and 1.44 for the silicon dioxide substrate. The dominant electric field of the fundamental transverse electric polarized mode (TE1) is in the radial direction, with an amplitude plotted in Fig.~\ref{Fig2}(b) with top view (the top panel) and cross-sectional view (the bottom panel). We find that the modes are well confined in this rod structure, and almost centered in the cell, even though the bending effect associated with the ring's radius of curvature is considered. As schematically illustrated in Fig.~\ref{Fig1}(a), the simulated $s0$ mode lies at the dielectric band edge and can be measured by cavity transmission spectroscopy. In Fig.~\ref{Fig2}(c), we clearly observe a large bandgap created in the spectrum, with no resonances observed between 1450~nm (the lower end of our wavelength scan range) and 1527~nm. The $s0$ mode is observed at 1527.2~nm (196.4~THz), and many other band-edge modes (identified by their reduced mode spacing relative to conventional WGMs far from the band-edge) are well coupled. A zoom-in of the band-edge modes (shaded in pale red and shown in the central panel of Fig.~\ref{Fig2}(c)) shows seven modes. These modes in general have high $Q$s around or above 10$^6$, with the $s0$ and $s1^+$ zoomed-in spectra shown in the rightmost two panels of Fig.~\ref{Fig2}(c). Infrared images (Fig.~\ref{Fig2}(d)) are taken with the focus plane at the microring surface and confirm the spatial patterns of these slow light modes around the circumference of the PhCR. The $s0$ mode slightly deviates from a uniform pattern, likely due to randomness in scattering. The $s1$ and $s2$ modes agree with the theoretical prediction of $|$cos($\phi$)$|$ and $|$cos(2$\phi$)$|$ intensity pattern ($\phi$ is the azimuthal angle), as illustrated in Fig.~\ref{Fig1}(b). We note that similar behavior has been observed for the MPhCR system~\cite{Lu_NatPhoton_2022_High}.

The rPhCR device geometry resembles a coupled-resonator optical waveguide, also termed a CROW~\cite{Yariv_OL_1999_Coupled, Mookherjea_JSTQE_2002_Coupled}, except that instead of a linear chain, the rod lattice is wrapped into a circular microring. The device also resembles a previous simulated work~\cite{Timmerman_OE_2022_Nanorod}, where the simulated $Q$ approaches $10^5$ for infinite-long circular rods, yet below $10^4$ for sub-micrometer rods. While the elliptical shape caters to WGMs better than the circular shape in our case, it is interesting to investigate the limitation of a circular rod PhCR, perhaps with a thicker film. The greater degree of symmetry of a circular rod unit cell can enable additional topological functionalities, particularly with respect to intersecting geometries. 

\begin{figure*}[t!]
\centering\includegraphics[width=0.88\linewidth]{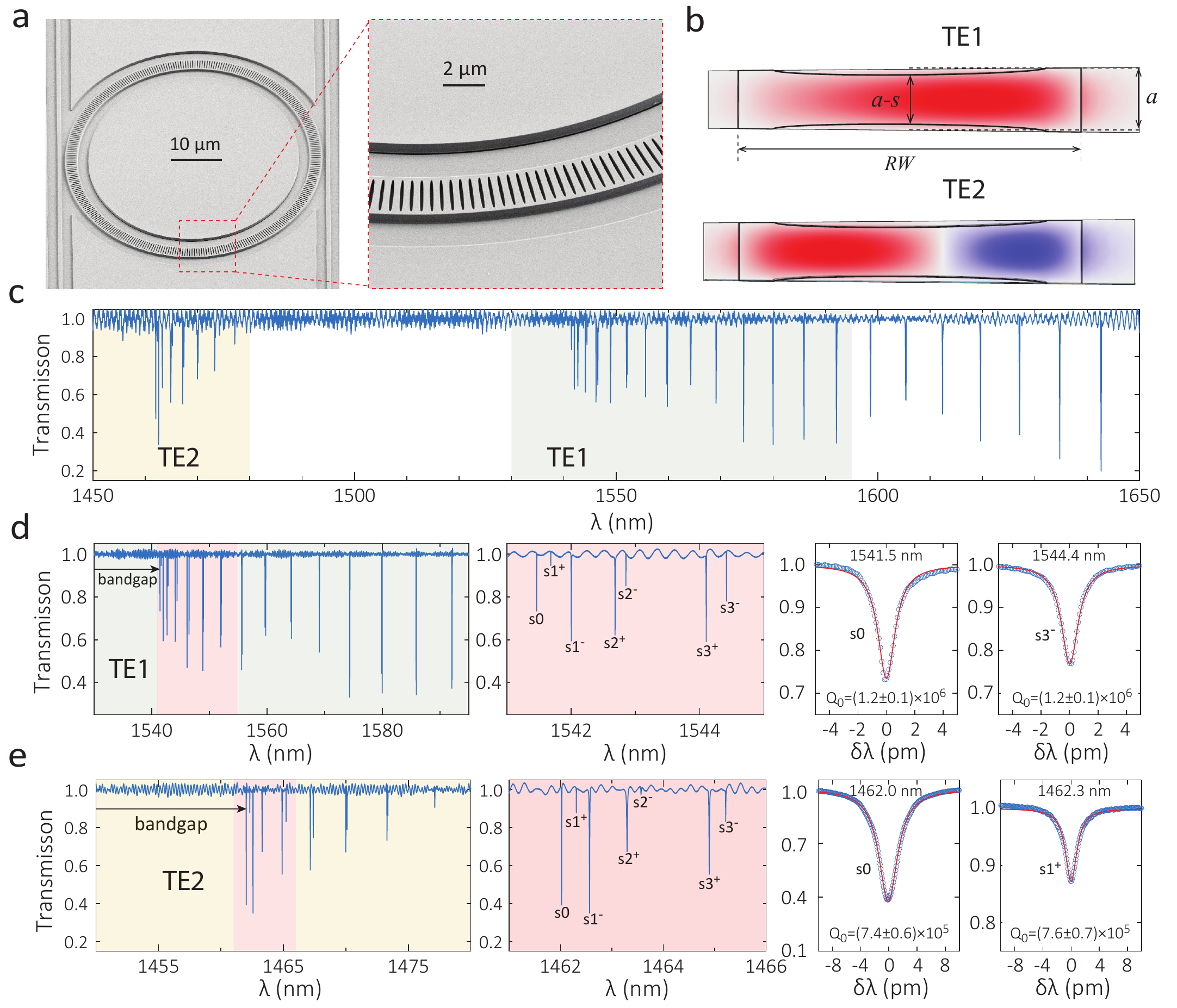}
\caption{\textbf{Introducing the `slit' photonic crystal ring (sPhCR).} \textbf{a,} SEM images of the sPhCR device and zoom-in view of the slit structure. The sPhCR has an outer ring radius of 25~$\mu$m, a ring width of 2.5~$\mu$m, and a thickness of $H =$~500~nm. There are 332 identical cells in this sPhCR with a period of $a =$~450~nm. Each cell has an elliptical air slit that is 2~$\mu$m long and $s =$~90~nm wide. \textbf{b,} A finite-element-method simulation of a unit cell of the sPhCR device, showing the dominant (radial direction) electric field profiles of the fundamental and 2nd-order band-edge modes in top view, labeled as TE1 and TE2, respectively. \textbf{c,} Measured transmission spectrum of the sPhCR device showing two sets of dielectric band modes. \textbf{d,e,} The modes at the TE1 and TE2 dielectric band-edge (pale red shaded areas) are zoomed-in and displayed in the middle panel, labeled as $\{s0,s1^{\pm},s2^{\pm},s3^{\pm}\}$. The right panels show nonlinear least squares fits to the $s0$ and $s3^-$/$s1^+$ modes, respectively. The uncertainty in $Q_0$ represents the 95~\% confidence range of the nonlinear fits.}
\label{Fig4}
\end{figure*}

\vspace{-0.3in}
\section{Defect localization}
\vspace{-0.2in}
\noindent Defect localization was previously reported with intuitive design in MPhCRs, where the defect is created by variation of a unit cell parameter (the PhC ring width modulation amplitude) across a number of cells, and leads to a localization of $s0$ mode to $g$ mode~\cite{Lu_NatPhoton_2022_High}. This is schematically illustrated in Fig.~\ref{Fig1}(a-c), where the frequency shift of the $g$ mode from the band edge is determined by the depth and width of the ring width modulation comprising the defect. In this section, we show that introducing defect localization in the rPhCR while maintaining high performance is also straightforward.

As shown in Fig.~\ref{Fig3}(a), we introduce a moderately-confined defect to the right part of the rPhCR. The defect is comprised of many cells (48 in the example shown here), with a quadratic variation in the rod linear dimensions ($RW$ and $rW$), with the central defect cell linear dimensions being 85~\% of $RW$ of an unperturbed unit cell. The center of the `rods' remains fixed in distance to the rPhCR center. We simulate a portion of the rPhCR containing the defect region using the finite-element method, and find the $g$ mode at approximately 198.1~THz. The mode profile is displayed in Fig.~\ref{Fig3}(b), indeed showing confinement within the modulated cells. The device transmission spectrum [Fig.~\ref{Fig3}(c)] shows the $g$ mode is located around 1536.39~nm (195.2~THz). The deviation from simulation is likely due to the deviation of the size of the defect cells from the targeted device pattern, and in principle can be calibrated if the geometry is characterized more accurately. We can see in Fig.~\ref{Fig3}(c) that the $g$ mode is deep in the bandgap, while six other modes are compressed around the band edge. While the $Q$ remains high (near 5$\times10^5$), the fit shows a decrease in $Q_0$ by a factor of 3 for the $g$ mode relative to the band-edge states of the rPhCR without defect, potentially due to parasitic loss associated with scattering induced by the access waveguide. The slow light modes have higher $Q_0$ values as they are under-coupled and free from this parasitic loss. The infrared images of the $s1$ and $s2$ band-edge modes in Fig.~\ref{Fig3}(d) are similar to those of the rPhCR without defect (Fig.~\ref{Fig2}(d)). In contrast to $s0$ in Fig.~\ref{Fig2}(d), the $g$ mode in Fig.~\ref{Fig3}(d) shows a clear localization within the defect cells (approximately 1/8 of the ring in dashed circle) with the central defect close to the waveguide (dashed vertical line) brightest.   

The increase in modal confinement from $s0$ to $g$ ($>$10$\times$) can be further optimized by incorporating deeper modulation across a smaller number of cells. We note that the rPhCR has more degree of freedoms in introducing the defect, in comparison to the MPhCR, as the boundary is no longer continuous. For example, the center of each rod can be shifted, and the orientation of each rod can be rotated. It is thus interesting to explore what combination of parameter modulation can lead to the highest $Q/V$ in design and fabrication. Beyond cQED, the rod geometry can be advantageous in nonlinear optical interaction in its confined volume and ideal mode overlap. For example, the localized defect mode can be promising in achieving efficient second harmonic generation~\cite{Lu_NatPhoton_2021_Efficient} with fundamental mode in the defect rPhCR mode ($g$) and second harmonic mode in a single nanorod~\cite{Koshelev_Science_2020_Subwavelength}. 

\vspace{-0.15in}
\section{`Slit' photonic crystal microring}
\vspace{-0.15in}
\noindent The success of the rPhCR with an elliptical unit cell, in contrast to the lower-$Q$ circular-rod PhCR previously studied, seems to suggest that a design principle for a high-$Q$ PhCR is to maximally cater (and minimally perturb) the conventional WGM mode shape. Following this principle, we try to shrink the aspect ratio of the air holes holes in previous PhCRs~\cite{Lee_OL_2012_Slow, Gao_SciRep_2016_Air, KML_OE_2014_Maximizing}, forming a 'slit' PhCR, or sPhCR. This operation is possible in PhCRs because only a bandgap along the propagation (azimuthal) direction is needed, in contrast to two-dimensional PhCRs, where a large-enough air-filling fraction is needed to open a full bandgap for all propagation directions.

We fabricate the sPhCR device and show SEM images in Fig.~\ref{Fig4}(a). These SEM images seem to indicate that the air slit is fully etched through the silicon nitride layer, but a full inspection (i.e., focused ion beam cut and SEM imaging) of its cross-section has not been performed. We carry out finite-element-method simulation of a unit cell using the parameters specified in the caption, and find two set of band-edge modes for the fundamental (TE1) and 2nd-order (TE2) mode families at 192.4~THz and 203.6~THz, respectively. The mode profiles are shown in top view, with TE2 having two colors (red and blue) indicating the the field node that results in the electrical field pointing in different radial directions (inward and outward). In the experimental device transmission spectrum of Fig.~\ref{Fig4}(c), these two bands are clearly observed, with TE1 and TE2 band edges at 1541.5~nm (194.6~THz) and 1462.0~nm (205.2~THz), respectively. The close correspondence of the simulations and experiments suggests that the slits are (or close to) fully etched. The TE1 modes show $Q_0$s above 10$^6$ and are mostly well coupled, while the TE2 modes show lower $Q_0=0.7\times$10$^6$, because of its larger overlap with the etched sidewalls and potential roughness.

The existence of two high-$Q$ mode families with intuitive design may be useful in cQED. For example, it is possible to have defect localization in these two mode families simultaneously, so that the TE2 defect mode can be used to AC Stark shift a quantum emitter into precise resonance with a TE1 defect wavelength. Through this method, high $Q/V$ can be supported simultaneously for both modes, resulting in potentially fast and efficient spectral control of the relative tuning of the coupled dipole-cavity system~\cite{bose_all-optical_2014}.

\vspace{-0.15in}
\section{Conclusions}
\vspace{-0.15in}
\noindent We report two types of photonic crystal microrings (PhCRs) with `rod' and `slit' unit cells, and show that high-$Q$ at the 10$^6$ level can be supported in both geometries. Taken together with previous work on microgear photonic crystal rings~\cite{Lu_NatPhoton_2022_High}, our work suggests that intuitive design for high-quality-factor PhCRs is in general straightforward, with a variety of different unit cell geometries possible. Combined with the ease of defect localization further verified in this work, and the capacity for multi-defect localization recently reported~\cite{Wang_PRL_2022_Fractional}, we believe that PhCRs are a promising platform for cavity quantum electrodynamics applications. Going forward, understanding both the design of high $Q/V$ air-mode PhCRs for interactions with gas phase atoms~\cite{Thompson_Science_2013_Coupling,Alaeian_APB_2020_Cavity} and extending our intuitive PhCR design approach to other photonic materials hosting different types of quantum emitters is of interest. We believe that the latter should in particular be straightforward to accomplish, assuming that smooth etching of the device sidewalls and reasonable preservation of the targeted pattern geometry can be achieved. 


\medskip
\noindent \textbf{Acknowledgement:}
The authors would like to thank Lu Chen and Yuncong Liu for helpful discussions.

\noindent \textbf{Author contributions:} X.L. and K.S. originated the idea. X.L. led the design, fabrication, measurements, and simulation. F.Z., Y.S., A.C., and M.W. assisted in the measurements. A.M., A.C., and M.D. assisted in the simulation. All authors participated in analysis. X.L., Q.Y., and K.S. wrote the manuscript, with the help from others. K.S. supervised the project.

\noindent \textbf{Research fundings:} This work is supported by the DARPA SAVaNT and NIST-on-a-chip programs, and partly sponsored by the Army Research Office under Cooperative Agreement Number W911NF-21-2-0106. M.W. is supported by the cooperative research agreement between University of Maryland and NIST, Award no. 70NANB10H193.

\noindent \textbf{Conflict of interest statement:} The authors declare no competing financial interests.

\bibliographystyle{ieeetr}

\end{document}